\documentclass[3p]{elsarticle}

\usepackage{color}
\usepackage{epstopdf}

\usepackage{amsmath}
\usepackage{graphicx}

\usepackage{amssymb}
\usepackage{hyperref} 

\biboptions{sort&compress}

\newcommand{\pd}{\partial}
\newcommand{\bs}{\boldsymbol}

\renewcommand{\d}{\mathrm{d}} 
\newcommand{\ii}{\mathrm{i}}
\let\oldpi\pi
\renewcommand{\pi}{\mathrm{\oldpi}}
\newcommand{\e}{\mathrm{e}}
\newcommand{\J}{\mathrm{J}}

\begin{document}

\begin{frontmatter}



\title{Shaping electron beams for the generation of innovative measurements in the (S)TEM}



\author[emat]{Jo Verbeeck}
\author[emat]{Giulio Guzzinati}
\author[emat]{Laura Clark}
\author[emat]{Roeland Juchtmans}
\author[emat]{Ruben Van Boxem}
\author[emat]{He Tian}
\author[emat]{Armand B\'ech\'e}
\author[emat,Triebenberglabor]{Axel Lubk}
\author[emat]{Gustaaf Van Tendeloo}

\address[emat]{EMAT, University of Antwerp, Groenenborgerlaan 171, 2020 Antwerp, Belgium}
\address[Triebenberglabor]{Triebenberglabor, University of Dresden, Zum Triebenberg 1, 01062 Dresden, Germany}

\begin{abstract}
In TEM, a typical goal consists of making a small electron probe in the sample plane in order to obtain high spatial resolution in scanning transmission electron microscopy. In order to do so, the phase of the electron wave is corrected to resemble a spherical wave compensating for aberrations in the magnetic lenses. In this contribution we discuss the advantage of changing the phase of an electron wave in a specific way in order to obtain fundamentally different electron probes opening up new application in the (S)TEM. We focus on electron vortex states as a specific family of waves with an azimuthal phase signature and discuss their properties, production and applications. The concepts presented here are rather general and also different classes of probes can be obtained in a similar fashion showing that electron probes can be tuned to optimise a specific measurement or interaction.
\end{abstract}


\end{frontmatter}


\section{Introduction}

It was recently predicted, and experimentally demonstrated, that changing the fundamental structure of electron wavefronts, can imbue them with a number of particular properties, enabling new functionality within a conventional transmission electron microscope (TEM)~\cite{Bliokh2007, Uchida2010, Verbeeck2010}.
These restructured electron beams, have continuous helical wavefronts, and are thus referred to as electron vortex beams~\cite{Uchida2010, Verbeeck2010}.
Accordingly the electron probability current follows a spiralling path, propagating along the optical axis, with an azimuthal momentum component~\cite{SchattschneiderTheoryOf}.

Vortices occur across a broad variety of physics subfields -- in classical fluid mechanics, superconductors, ocean tides and collective human behaviour~\cite{saffmanvortex, superconductorvortices, Whewell,  nyetides, moshpitvortex}.
Screw--type phase singularities were first discussed as a ubiquitous feature of wave physics in 1974~\cite{Nye1974}, followed by studies progressing into the deliberate production of optical beams with the azimuthal phase dependence, in the work of Vaughan and Willetts~\cite{VaughanWilletts} and Bazhenov \emph{et al. }~\cite{bazhenovscrew}.

A simple vortex beam can be described by $\Psi\propto f(r)\exp(\ii \ell \varphi)\exp(\ii k_z z)$ where $(r, \varphi, z)$ are the cylindrical coordinates, and $k_z$ is the forward momentum of the beam.
The order of the vortex is $\ell$, wherein the phase changes by $2\pi\ell$ in a circuit around the vortex core.
In this single vortex case, $\ell$ is equal to the winding number \cite{lubk2013}, a key parameter describing the field vorticity. 
The vortex order is also related to the orbital angular momentum (OAM) of the beam, a property that has been of much interest in the field of singular optics \cite{Allen1992, OptAngMomBook, BabikerNewBook}.

\begin{figure}
\begin{center}
\includegraphics[width=\linewidth]{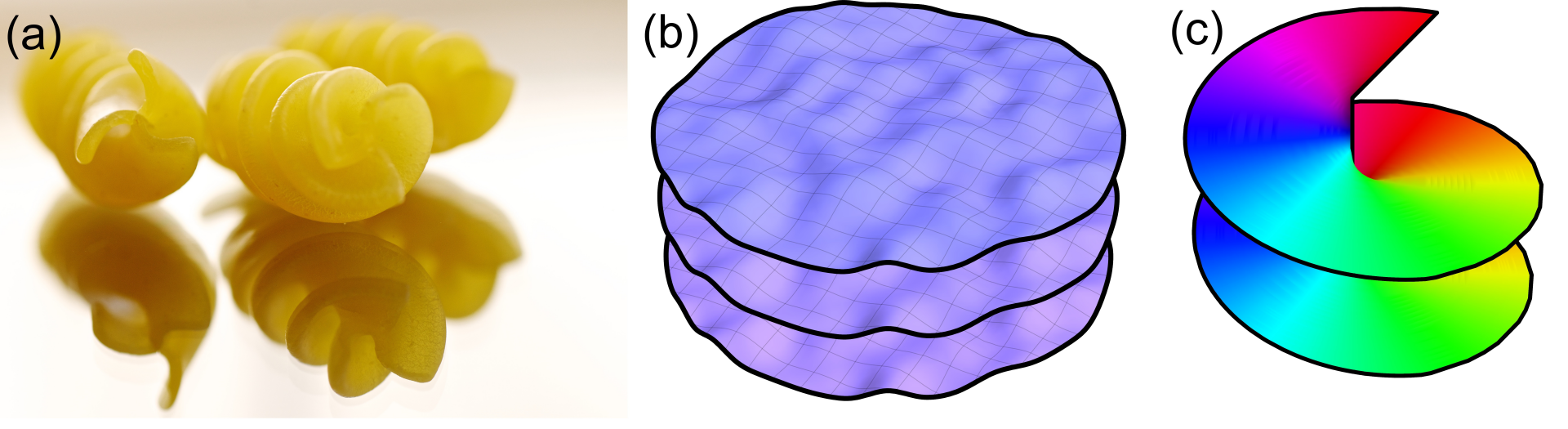}
\end{center}
\caption{a) A common example of the $\ell=3$ wavefront structure, demonstrated in fusilli pasta~\cite{fusilliphoto}. b) Equiphase surfaces of a plane wave with random smooth distortions. c) Equiphase surface of an $\ell=1$ vortex beam. \label{pastafig}}
\end{figure}

To visualise such beams easily, the wavefronts of an $\ell=\pm 3$ vortex beam are of the same structure as a piece of fusilli pasta, as illustrated in figure \ref{pastafig}a.

Along the axis of a vortex beam, all phases meet, and as such there is a indeterminacy in the phase along the axis, called a phase singularity, or in 3D, a vortex line \cite{berry2001knotted, dennisthesis}.
The screw structured phase singularity was first discussed by Nye and Berry in 1974~\cite{Nye1974}.
As a consequence of the meeting of phases, the intensity on axis is forced to zero by destructive interference.
Indeed, this is a key indicator of a vortex beam, accessible without phase measurement techniques; the central zero--intensity of a vortex beam remains present throughout propagation, while in a non--vortex, hollow beam, the central minimum will disappear over propagation due to diffractive spreading.
A further result of the phase singularity is that a vortex beam is topologically distinct from a planar or spherical wavefront -- no continuous stretching or warping can cause the creation of a vortex wavefront from planar or spherical surfaces, as illustrated in figure \ref{pastafig}~\cite{lubk2013}.

Following the development of vortex beam theory throughout the later part of the $20^{th}$ century, vortex beams of photons, electrons, and acoustic waves, have now found practical applications in a broad range of fields. They have significant applications in atomic and molecular manipulation, trapping and rotation, including work with Bose--Einstein condensates \cite{andrewsbook, Verbeeck2013}.
They have been demonstrated across a broad range of the electromagnetic spectrum \cite{xrayvortex, radiovortex, irvortex, uvvortex}.
Acoustic vortex beams have been studied by a number of groups \cite{skeldonacoustic, torabiacoustic}, while optical vortex beams are finding further application beyond typical optics studies in astrophysics \cite{berkhoutastro}, and in telecommunications \cite{wangterabit, vortexnot, andrewsbook}.

\section{Vortex beam properties}

\begin{figure}
\begin{center}
\includegraphics[width=.9\linewidth]{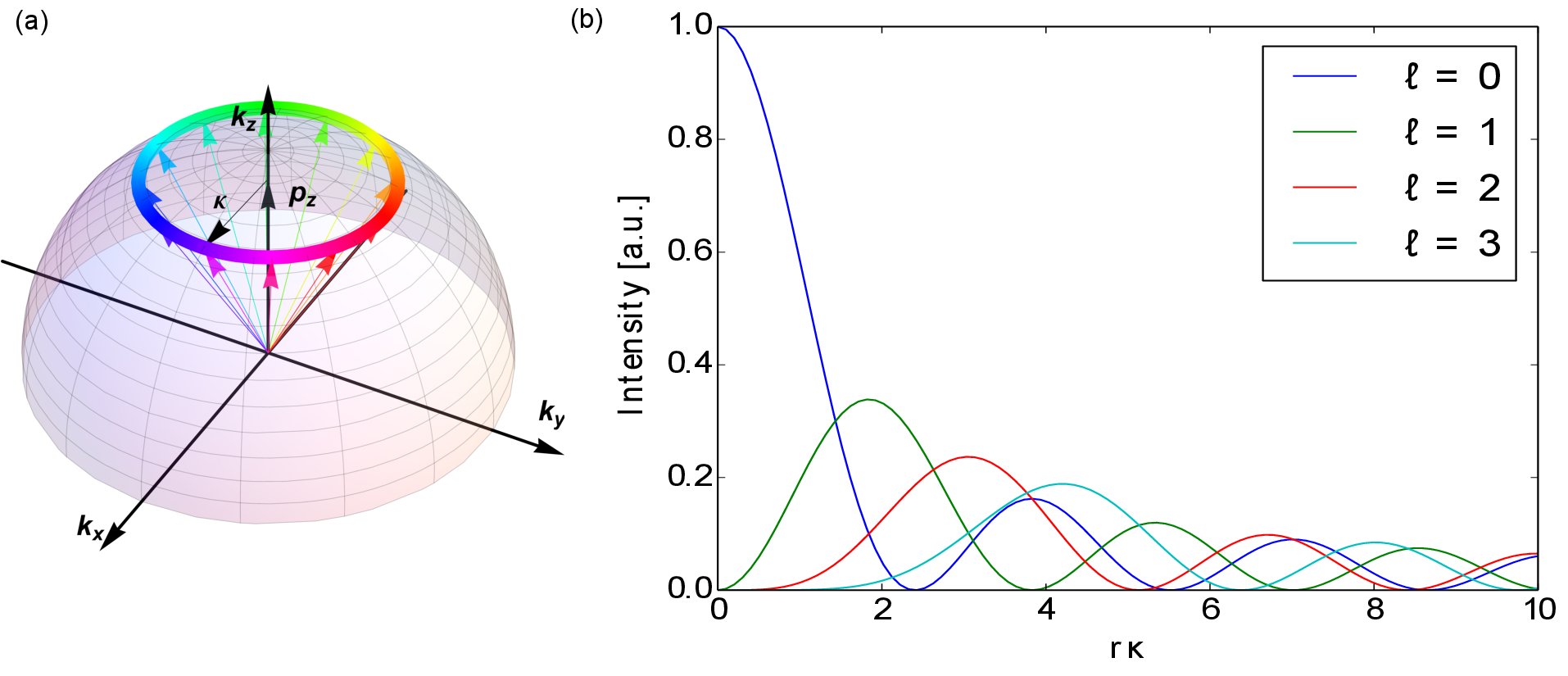}
\end{center}
\caption{(a) Fourier space representation of a Bessel beam $\Psi_1$ (see eq.~\eqref{eq:bessel_beam}). The radius of the ring is equal to the transverse momentum $k_\perp$. (b) Radial intensity for different topological charge. For $l=0$ we obtain the well known Airy disc, for $l\neq 0$ doughnut like intensity distributions are obtained. \label{fig:bessel_fourier_intensity}}
\end{figure}

The simplest scalar wave function solution describing particles with orbital angular momentum (OAM) is the \emph{Bessel beam}~\cite{Durnin_bessel_beams},
\begin{equation} \label{eq:bessel_beam}
\Psi_\ell(r,\varphi,z) = \e^{\ii k_z z} \e^{\ii \ell\varphi} \J_\ell(k_\perp r),
\end{equation}
which is an exact solution of the time-independent Schr\"odinger equation and an eigenvector of the cylindrical free space Hamiltonian, the longitudinal momentum $p_z$, the transverse momentum $p_\perp = \ii \hbar \sqrt{\pd_x^2+\pd_y^2}$, and the OAM $L_z~=~-\ii\hbar\pd_\varphi$.
Here, $\hbar\ell$ is the OAM eigenvalue, $\J_\ell$ is the $\ell$-th cylindrical Bessel function of the first kind, and $k_\perp$ is the transverse momentum of the vortex state.
The transverse momentum $k_\perp$ is the eigenvalue of the operator $p_\perp$.
Figure \ref{fig:bessel_fourier_intensity}.a shows the Fourier transform of a Bessel beam, which is, up to normalization, a Dirac delta ring multiplied by a phase factor.
Bessel beams are the formal basis functions of the cylindrical Schr\"odinger equation, and a single state as in eq.~\eqref{eq:bessel_beam} cannot be realized in experiments much the same as a true plane wave can not exist in experiments.
The Bessel beam intensity profiles (modulus squared of the wave function in eq.~\eqref{eq:bessel_beam}) for several $\ell$ are shown in figure \ref{fig:bessel_fourier_intensity}b.
A circular aperture with spiral phase structure contains a single $\ell$ mode Bessel beam superposition of the following form:
\begin{equation}
\Psi \propto \e^{\ii\ell\varphi} \int_0^{k_\perp^\mathrm{max}} \d k_\perp k_\perp \J_\ell(k_\perp r),
\end{equation}
containing a sum of $k_\perp$ eigenstates up to a maximum $k_\perp^\mathrm{max}$ proportional to the radius of the aperture. For $\ell=0$ we recover the typical Airy disc while for other values of $\ell$ a slightly broader vortex beam is obtained for the same $k_\perp^\mathrm{max}$.
The smallest electron vortex beam diameter obtained thus far had a diameter of 1.2~\AA~\cite{Verbeeck_nanoresearch}.

The above solutions of the Schr\"odinger equation will change significantly in a magnetic field, and depending on the exact form of the external field, propagation dynamics and angular momentum, will change~\cite{Bliokh2012e}.
Fully relativistic solutions have been investigated as well by several authors, but the free space relativistic effects have been found to be negligible~\cite{Bliokh_spin_orbit,VanBoxem2013}.



A scalar electron vortex possesses a topological charge defined by its winding number 
\begin{equation} \label{eq:winding_number}
\ell = \frac{1}{2\pi} \oint \d\phi = \frac{1}{2\pi} \oint \frac{\nabla \Psi}{\Psi} \cdot \d\boldsymbol s.
\end{equation}
This distinguishes a vortex wave from a plane wave.
Lubk \textit{et al.} showed that the propagation of an electron vortex is \emph{topologically protected}, \textit{i.e.} a vortex core will not delocalize as rapidly as a conventional $\ell=0$ probe~\cite{lubk2013,Xin}.
One must be careful relating the winding number in eq.~\eqref{eq:winding_number} directly to the OAM as they are only equal in the case of an OAM eigenstate, which means the electron density is radially symmetric around the phase dislocation.
For better clarity, we will suppose in the following that this condition is always fulfilled.


Vortex beam states as discussed in eq.~\eqref{eq:bessel_beam} are eigenvectors of the OAM operator for which:
\begin{equation}
\langle L_z \rangle = \hbar \ell.
\end{equation}
The current experimental record for electron vortices' OAM is $\ell=$100, although this was in the weak higher diffraction orders of a holographic fork aperture~\cite{McMorran2011}.
Translating and rotating the OAM axis with respect to the symmetry axis results in a change in the expectation value of $L_z$ and the decomposition in its components~\cite{berryOpticalCurrents,berkhoutthesis,Vasnetsov}.
In scattering experiments, this is an important detail to take into account~\cite{Loffler_crystal,lubk2013,Verbeeck_nanoresearch}.
Additionally, a scattered electron wave may have its principal OAM axis rotated by the scattering event, in which case further care needs to be taken to calculate and measure the outgoing electron wave~\cite{Ivanov_orbital_helicity}.


Bessel beams as in eq.~\eqref{eq:bessel_beam} show a spiralling current approximately equal to~\cite{Bliokh2007}:
\begin{equation}
\boldsymbol j(r) \approx \rho(r) \left( k_z \boldsymbol e_z + \frac{\hbar \ell}{r}\boldsymbol e_\varphi \right).
\end{equation}
This current lies perpendicular to the (spiralling) wave front's equiphase surface.
Figures \ref{pastafig}b-c compare a continuously distorted wave front's equiphase surface, and that of a vortex beam.
It is immediately evident that one cannot be simply transformed into the other without introducing a discontinuity in both the current and the phase surface.


The electromagnetic fields of a Bessel beam are given by~\cite{Lloyd_fields}:
\begin{subequations}
\begin{align}
\boldsymbol E(r) &\propto  - r \left[\J_\ell^2 (k_\perp r) - \J_{\ell-1}(k_\perp r)\J_{\ell+1}(k_\perp r) \right] \boldsymbol e_r \\
\boldsymbol B &= B_\varphi \boldsymbol e_\varphi + B_z \boldsymbol e_z \\
B_\varphi &\propto r \left[ \J_\ell^2 (k_\perp r) - \J_{\ell-1}(k_\perp r) \J_{\ell+1}(k_\perp r) \right] \\
B_z &\propto \frac{\ell}{r} \int_r^\infty \J_\ell^2(k_\perp r^\prime)\d r^\prime
\end{align}
\end{subequations}
The magnetic field in the $z$ direction, $B_z$, peaks at the center of the vortex.
Note that considering a more realistic wave packet will add fringe fields which do not appear in the unlocalized and unnormalized Bessel beam case.
It will also remedy the fact that in the above expression, $B_z$ is always positive and there are no returning field lines.


Another interesting feature of electron vortex states (as in eq.~\eqref{eq:bessel_beam}) is their magnetic moment, $\bs \mu$, which is proportional to the OAM~\cite{Bliokh2012e}:
\begin{equation}
\bs \mu = \frac{e \hbar}{2} \frac{\int \boldsymbol r \times \boldsymbol j \d V}{\int \rho \d V} = \ell \mu_B \bs e_z.
\end{equation}
This can be arbitrarily tuned by changing the OAM of the beam.

\section{Vortex creation}

A variety of methods, many derived from optics, have been used to create electron vortex beams. The purpose of these techniques is the controlled manipulation of the phase of the electron wave in order to generate the characteristic singularity. It should be noted that these methods are often entirely general, and can be used to obtain many other types of wave, such as electron Airy beams which have recently gained interest \cite{Voloch-Bloch2013}.


Holographic reconstruction has so far been the most common method for the production of electron vortex beams. In conventional electron holography one interferes the wave transmitted through a sample with a reference wave, creating a hologram that encodes the phase and amplitude in the interference pattern. In holographic reconstruction one makes use of a computer generated hologram in order to reproduce a target wave with a desired amplitude and phase profile. The hologram is calculated by superimposing the target function ($\Psi \propto \exp{(\ii \ell \varphi)} $ for a vortex beam) with a reference wave such as a tilted plane wave of the form $\exp{\ii k_x x}$. The hologram is binarised by using a threshold which is needed as a true amplitude modulation in TEM is very difficult to realise. The binary hologram can then be manufactured through standard nanofabrication techniques such as electron beam lithography (EBL), electron beam induced deposition (EBID) or most commonly milled with a focused ion beam (FIB) instrument on a metallic thin film of heavy, strongly scattering metals such as platinum or gold \cite{Verbeeck2010,McMorran2011}.

\begin{figure}
  \begin{center}
    \includegraphics[width=.9\linewidth]{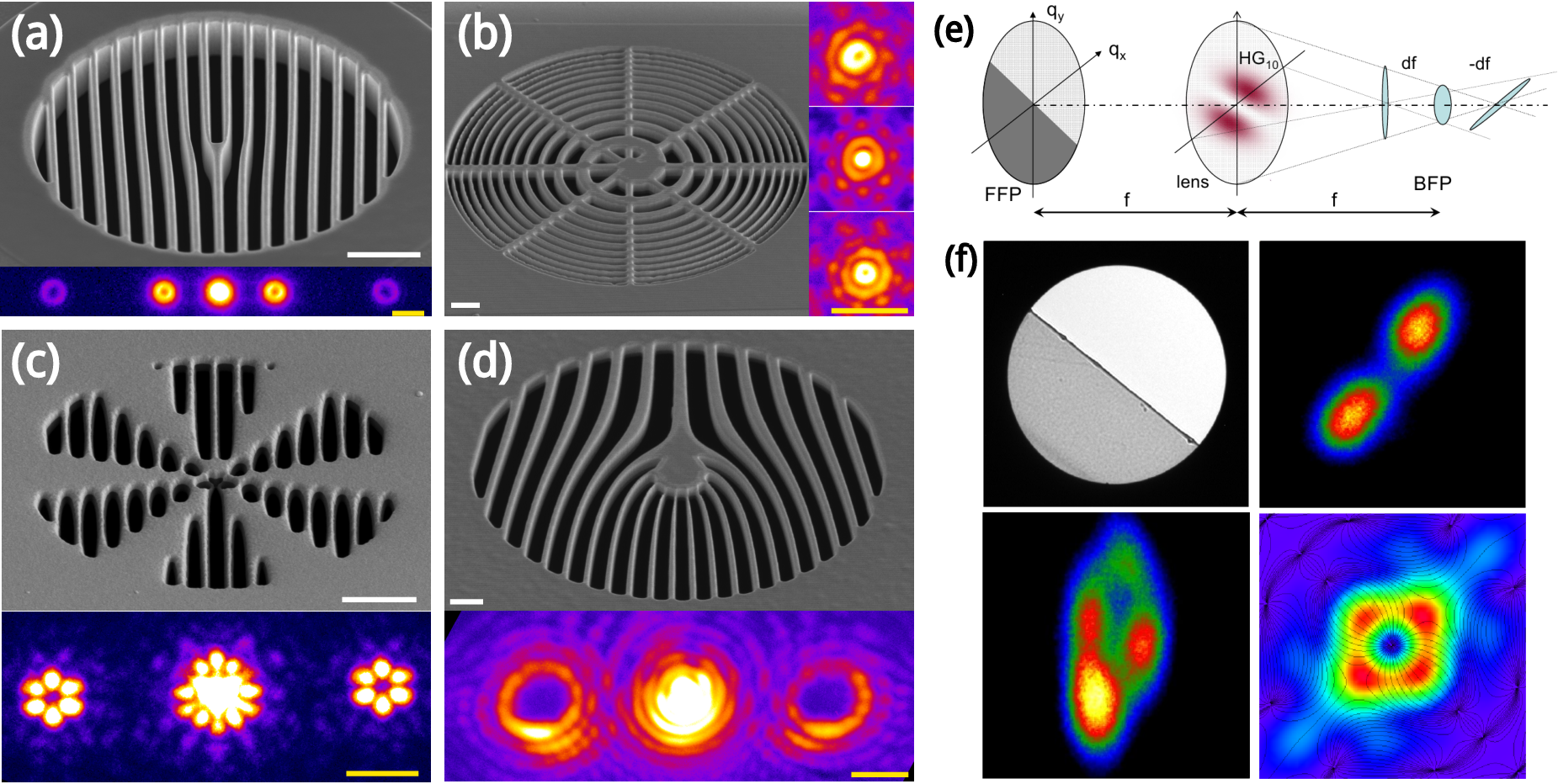}
    \caption{(color online) Examples of holographic reconstruction: SEM micrographs of the holograms are shown alongside the beams they produce when used in a TEM. The hologram images have a 2~$\mu$m white scalebar, the beam intensities have 20~nm yellow scalebar.
    (a) Example of $\ell=1$ fork hologram. The $\ell=0$ beam is visible in the center, with the first sidebands of $\ell=\pm 1$ and the higher order with $\ell=\pm 3$ at the border of the image.
    (b) Spiral hologram generated by the interference of an $\ell=1$ vortex beam with a spherical wave. The probes shown on the side are separated along the beam axis.
    (c) The target wave used was the flower shaped superpositions between $\ell=3$ and $\ell=-3$ vortex modes.
    (d) Higher order $\ell=\pm 9$ vortex modes.\label{fig:holography}
    (e) Diagram of the vortex generation through mode conversion. The Hilbert phase plate in the front focal plane of a lens introduces a phase shif of $\pi$ in the incident plane wave, generating a beam with two lobes that is then converted into a vortex beam by introducing astigmatism
    (f) Top: Shadow image of the phase plate and resulting beam produced by the phase shift. Bottom: Experimental generated vortex beam and simulation of the ideal result. Despite the distortions, attributed to the partial absorption by the phase plate, the central phase singularity is clearly visible.
    (e) and (f) from Schattschneider \emph{et al.} 2012 \cite{Schattschneider2012a}}
  \end{center}
\end{figure}

When such a hologram is inserted in the TEM and illuminated with a sufficiently coherent electron beam the desired vortex beam can be observed in the far--field of the mask hologram. The diffraction pattern contains a central spot from the reference wave and two sidebands with one having the desired waveform and the other being its complex conjugate (see Fig. \ref{fig:holography}). The binarization of the mask introduces higher diffraction orders with higher OAM that can be seen at higher distance from the center of the pattern. The ensemble of these diffracted beams forms a one dimensional vortex lattice (see Fig. \ref{fig:holography}.a).
This method allows for several variations by changing either the reference or the target wave. If a spherical reference wave is used instead of a plane wave the generated hologram assumes a spiral shape and the various vortex orders are separated longitudinally along the beam axis as in Fig \ref{fig:holography}.b. This approach is especially suited for scanning transmission electron microscopy (STEM) since one vortex order can be focused on the sample at a given time and therefore give spatial information, while the other vortex probes are out of focus and provide only a background signal \cite{Verbeeck2011a}. Atomically resolved STEM images with vortex probes were obtained this way\cite{Verbeeck2011a}.
If the target wave is changed, the holographic method allows the generation of superpositions of vortex beams \cite{Greenshields2012,Guzzinati2013} (Fig. \ref{fig:holography}.c), beams with higher order topological charge \cite{McMorran2011,Saitoh2012a} (Fig. \ref{fig:holography}.d)
and electron Airy beams \cite{Voloch-Bloch2013}. This last type of beam is of note, as the target phase is completely different from that of a vortex and its feasibility demonstrates the flexibility of holographic reconstruction.

The phase of an electron beam is also influenced by electric and magnetic fields, such as the mean inner potential of a thin sample which induces a phase shift in an electron wave traveling through it.
A weak potential $V_0$ will induce a phase shift on the electron beam of $\Delta \phi = \frac{\pi}{2 \lambda E} \int V_0 ( \vec r )dz$ where $\lambda$ is the electron wavelength and $E$ is its energy. This has been exploited in the  fabrication of phase plates for electron microscopy with the purpose of enhancing the contrast in weakly scattering samples \cite{Danev2001,Danev2002}.
Examples of such phase plates are Zernike phase plates \cite{Zernike1942,Zernike1942a}, which induce a phase shift of $ \pi/2 $ in scattered waves for better phase contrast and are realized by a thin carbon film with a hole drilled in the center \cite{Danev2001},
or Hilbert phase plates, which produce a topological contrast by inducing a phase shift of $\pi$ between the lobes of a beam and are made of a semicircular thin film \cite{Danev2002}. Electrostatic Boersch phase plates exist as well creating a phase shift with a microfabricated electrostatic lens \cite{Schultheiss2006}

In order to produce a vortex wave, a spiral phase plate is needed and can in principle be produced using the same principles as the simpler phase plates discussed above. An approximate spiral phase plate, given by a spontaneous staircase arrangement of graphite thin films in a sample, was indeed used in the first demonstration that electron vortex beams do occur in a TEM \cite{Uchida2010}.

However such phase plate is extremely challenging to produce. If made out of carbon and designed for an electron beam with a kinetic energy of 300 keV the required thickness would be about 80~nm in the thickest point and should then be stepped gradually down to 0~nm. This degree of thickness control is extremely hard to reach with current technology, and the resulting object would suffer from a higher degree of contamination and radiation damage as compared to the binary holograms \cite{Danev2001}. Using heavier elements would require an even lower thickness and scattering would further complicate the electron interaction. It is worth noting that these phase plates have to be tuned for a specific acceleration voltage of the TEM, as the acquired phase shift depends strongly on it.

Hilbert phase plates have also been used in combination with a controlled level of astigmatism to produce electron vortices \cite{Schattschneider2012a},
through the so--called \emph{mode conversion} process \cite{Allen1992,Beijersbergen1993} (figure \ref{fig:holography}.e,f).


Despite the challenges presented in producing an electron phase plate, direct manipulation of the phase is performed when correcting the aberration of electron lenses. In fact, aberration correctors are designed to generate a phase plate to compensate for the aberration phase shifts of a lens system but also allow the free tuning of the individual aberrations to obtain a desired phase plate. Indeed while a high degree of aberration has been shown to induce a diffraction catastrophe that generates a vortex lattice \cite{Petersen2013},
it has been shown that a careful tuning of different orders of astigmatism, together with an angle-limiting annular aperture, can be used to produce a single high intensity vortex beam \cite{Clark2013}.

Magnetic fields also induce a phase shift in an electron wave through the Aharonov-Bohm effect. Indeed two different electron paths are phase shifted with respect to each other by $\Delta \phi = \oint \bs A \cdot \d \bs s = \int_{\Sigma} \bs B \cdot \d \bs S$.
The Aharonov-Bohm phase shift is commonly exploited through electron holography to measure the in-plane component of the magnetization \cite{Tonomura1982}.
The fabrication of magnetic phase plates for contrast enhancement through the Aharonov-Bohm phase is also currently under study \cite{Edgcombe2012}. It has been shown that there is an intimate link between the Aharonov-Bohm effect and electron vortex beams \cite{Bliokh2012e}
and that the characteristic field of a magnetic monopole would introduce an azimuthal phase shift that would turn a plane wave into a vortex with an OAM eigenvalue equal to the monopole's magnetic charge \cite{Tonomura1987,Bliokh2007}.
Recently it has been shown how a magnetic monopole field can be approximated by the end of a thin magnetic needle, providing a vortex beam generation mechanism that works independently of the electron kinetic energy. This method can produce electron vortex beams with high current and potentially very high OAM.

After this discussion of the techniques devised for the production of vortex beams, it is worth noting that electron vortices are also created in the inelastic collisions between the incident electrons and the atoms of a sample. The exchange of angular momentum in the interaction puts the inelastically scattered electron in a mixed state that naturally contains vortices \cite{Schattschneider2012}. In that sense, electron vortices have always existed and are nearly impossible to avoid.

\section{Applications}

Electron vortex beams have great potential in a number of applications, some of which have been demonstrated experimentally and others which thus far remain theoretical concepts.
\subsection{Nanomanipulation}
\begin{figure}
  \begin{center}
    \includegraphics[width=1\linewidth]{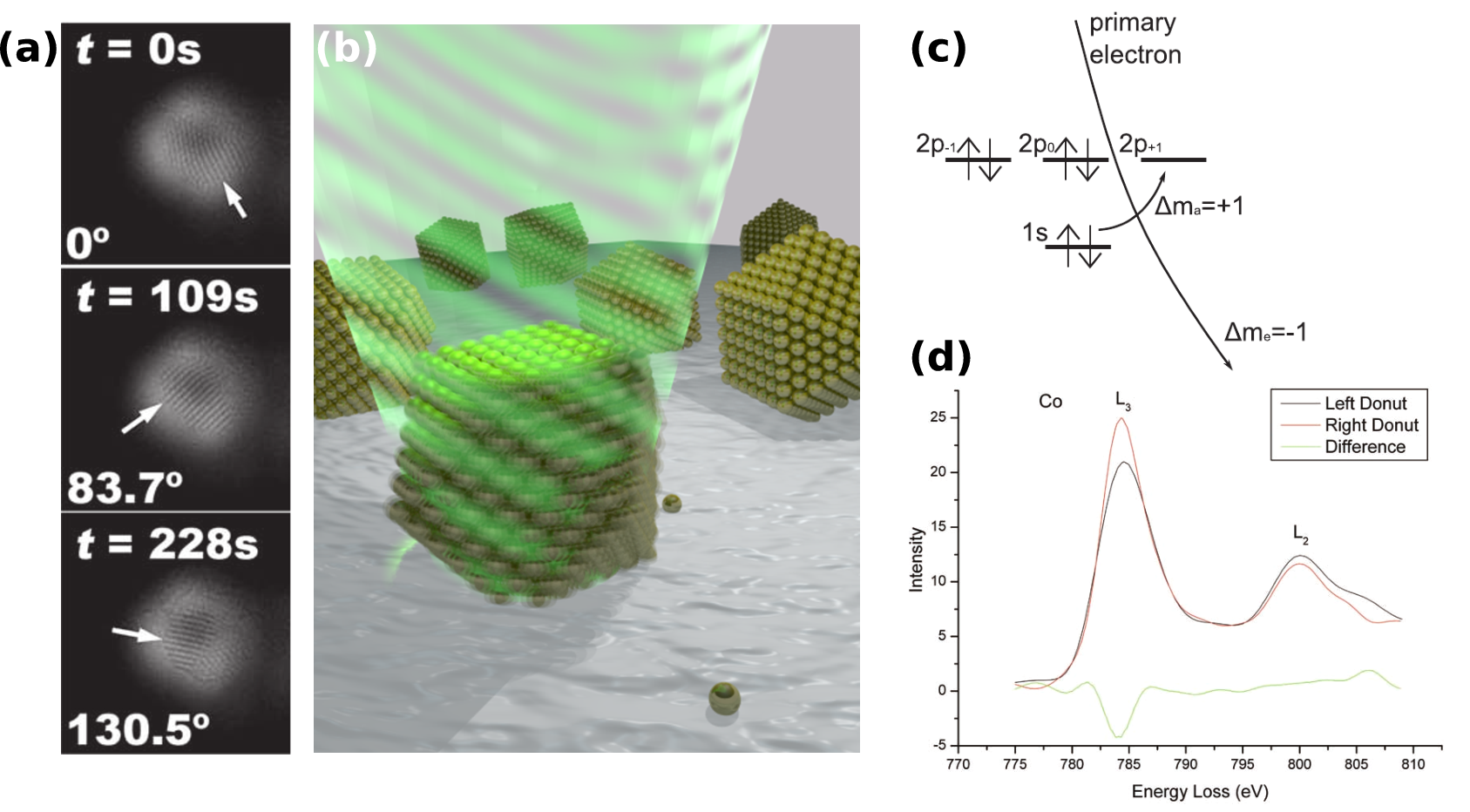}
    \caption{(a)Time series of atomic resolution images of a  3~nm gold particle illuminated by an $\ell=1$  vortex beam. The arrow denotes the direction of the (111) planes. From Verbeeck \emph{et al.} 2013.\label{figVerbeeck}
(b) Artistic representation of a gold particle illuminated by a vortex beam.
(c) Scheme of the electronic transition under consideration, $1s\rightarrow2p_{+1}$ showing a different cross section for the $\Delta m_1$ transition when illuminated with a $\ell=+1$ or a $\ell=-1$ beam.
(d) Measured EELS signal from left and the right sideband showing a clear difference on the $L_{2,3}$ edges. From Verbeeck \emph{et al.} 2013. \label{EMCDSignal}}
  \end{center}
\end{figure}

As stated in the previous section, electron vortex beams carry a well--defined orbital angular momentum of $\ell\hbar$ per electron. When scattered at a particle though, radial symmetry is broken and OAM is no longer a good quantum number \cite{Loffler_crystal}. Consequently transfer of OAM between the vortex and the particle is possible \cite{Verbeeck2013}.
\\
This effect is demonstrated by Verbeeck \emph{ et al.} \cite{Verbeeck2013}, where a gold nanoparticle with a diameter of 3~nm is illuminated by an $\ell=\pm1$ vortex of comparable size. By looking at the (111) planes of the gold particle in time, a rotation of $0.01$~rad/s is measured, as visible in fig. \ref{figVerbeeck}a. The sense of this rotation is determined by the sign of $\ell$ demonstrating that the rotation of the gold particle is a consequence of the vortex character of the electrons. It was calculated numerically an OAM of about  $0.1\hbar$ per electron is transferred to the crystal. The size of this transfer depends crucially on the orientation of the particle with respect to the beam. Neglecting friction, this would result in a linearly increasing rotational velocity of approximately $10^{14}$~rad/s$^{2}$ for a typical beam current of 50~pA which would cause the particle to disintegrate very rapidly. Incorporating friction however, a stationary rotation velocity of 0.037~rad/s is calculated, much closer to the observed value.\\
This experiment clearly shows that electrons carrying OAM can be used to rotate nanoparticles and could be a useful complement of atomic force microscopy techniques for manipulating nanoparticles. In addition electron vortex beams offer a new tool for studying rotational friction of nanoparticles \cite{Verbeeck2013}.

\subsection{Energy--loss magnetic circular dichroism (EMCD)}
Dichroism is the effect in which the photon absorption cross-section spectrum of a material depends on the polarization of the incoming photon. In X-ray magnetic circular dichroism (XMCD), first observed on the K--edge signal of iron \cite{Schutz1987}, the absorption cross-section depends on the angle between the helicity of a circularly polarized photon and the magnetization of a ferromagnetic or paramagnetic material. By looking at the energy loss spectrum of the photons, information about the magnetic ordering of the material can be obtained.  Recently Schattschneider \emph{et al.} demonstrated that the same can be done using electrons (EMCD) by exploiting the equivalence between the polarization vector $\bs \epsilon$ in X-ray absorption and the momentum transfer $\hbar\bs q$ in electron scattering~\cite{Schattschneider2006}.\\
Verbeeck \emph{et al.} suggest  another EMCD setup for mapping the spin-polarized $2p\rightarrow3d$ electronic transitions from a thin homogeneously magnetized ferromagnetic iron sample~\cite{Verbeeck2010}. These transitions can transfer OAM to an inelastically scattered electron if an asymmetry between the inelastic transition of $\Delta \ell=1$ and $\Delta \ell=-1$ exists. This asymmetry is always present in atoms carrying a magnetic moment mediated by a complex spin orbit coupling of the atomic states\cite{Schattschneider2010_with_Ennen}. As a consequence, the 
inelastic electrons carry a net excess of OAM with a different sign for the $L_2$ and $L_3$ excitations in electron energy loss spectroscopy. 
This imbalance in OAM  versus energy loss is measured by placing an $\ell=1$ fork hologram in the selected area plane and then placing the spectrometer entrance aperture in the center of the sidebands. The difference in the spectra obtained in the two sidebands is then due only to the magnetic transitions as shown in fig. \ref{figVerbeeck}d, much in the same way as in XMCD. The advantage of this method over XMCD is however that atomic resolution is feasible with electron vortices. Preliminary related experiments indicate that this indeed seems possible~\cite{Schattschneider_spin_polarized} although a sub-optimal signal to noise ratio makes the interpretation difficult.

\subsection{Vortex electron energy loss spectroscopy for near-field mapping of magnetic plasmons}
When passing through or moving in the vicinity of a material, an electron can excite plasmons in a material. Recording the energy lost in this process by means of electron energy loss spectroscopy (EELS), a map of the electrical part of the local plasmonic response of nano-materials can be reconstructed \cite{Pettit1975,Nelayah2007}.
One could expect that a vortex electron, carrying a magnetic moment of $\ell\mu_B$, could be used to probe the magnetic part of the local plasmonic response of a particle down to the nanometer scale.\\
The magnetic response of a split-ring resonator (SRR) is calculated using a semiclassical approach in Mohammadi \emph{et al.} \cite{Mohammadi2012}. Conventional finite-difference time domain (FDTD) calculations were performed for calculating normal EELS scattering probabilities in which the electric charge is replaced by an effective magnetic charge. It was shown that the magnetic component of the EELS signal is of the same order as the conventional EELS signal. This opens up the route for using vortex beams as a tool for mapping the magnetic response of nano-particles and could be of great importance in the search for artificial metal nanostructures with a large magnetic response in the visible light spectrum. Such nanostructures could be applicable in, for example, perfect lenses and optical cloaking \cite{Smith2004,Alu2007}.

\subsection{Spin-Orbital conversion}

In the following a short outline is given of how a space-variant Wien filter can be used as a spin-polarizer when using electron vortices following Karimi \emph{et al.} \cite{Karimi2012}.
Considering a spin unpolarized vortex-beam moving along the $z$-axis with topological charge $\ell$, $\Psi=\frac{1}{\sqrt{2}}\left(|\uparrow,\ell\rangle + |\downarrow,\ell\rangle\right)$, where the spin is taken along the propagation axis. When moving through a magnetic field of the form $\bs B(r,\phi,z)=B_0(r)(\cos(\alpha (\phi)),\sin(\alpha(\phi)),0)$, with $\alpha(\phi)=m\phi+\beta$ and $\phi$, being the azimuthal coordinate, part of the spin-up component will flip to a spin-down state and vice versa. Depending on the initial spin-state, the spin-flipped part of the wave function will gain or lose an amount of OAM equal to $m\hbar$, that is \cite{Karimi2012}:
\begin{align}
&|\uparrow,\ell \rangle \rightarrow \cos (\delta/2)|\uparrow,\ell\rangle + \ii \e^{\ii\beta} \sin(\delta/2) |\downarrow,\ell+m\rangle,\nonumber\\
&|\downarrow,\ell\rangle\rightarrow \cos (\delta/2) |\downarrow,\ell\rangle+\ii \e^{-\ii\beta}\sin(\delta/2)|\uparrow,\ell-m\rangle, \label{Filter}
\end{align}
where $\delta=gL/R_c$ and $R_c$ is the cyclotron radius. It can be seen from eq. \eqref{Filter} that when $\ell=\pm m$, part of the vortex beam will lose all of its OAM and that this part will be spin-polarized. When propagating to the far--field, the $\ell=0$ component of the electron wave will cause a finite amount of intensity to be present in the center of the beam, while the OAM containing components will not. By placing a small aperture in this plane, the $\ell=0$ part of the beam can be selected, resulting in a spin-polarized beam.\\
According to Karimi \emph{et al.}, the beam can have a degree of polarization up to 97.5\%, while maintaining an intensity of two orders of magnitude greater than conventional spin-polarised electron sources \cite{Karimi2012}.
Although exchange effects are seen to be negligible in current TEM experiments, a high-intensity polarized electron source would be a breakthrough~\cite{lubk2013}.

\section{Outlook and conclusion}

Electron waves provide more flexibility in amplitude and phase than the small subset of plane wave and spherical wave which are commonly used in a TEM. In this review we have shown how the new class of vortex waves with an azimuthal phase signature can expand the possibilities in TEM beyond the current state of the art. We have shown the peculiar and specific properties of these vortex beams as well as a wide range of methods by which they can be produced. As the creation of these waves is relatively new, their applications are just starting to emerge. The most promising application is undoubtedly the appearance of magnetic information in EELS which offers the potential of probing magnetic moments on the atomic scale. 
Compared to optical vortices which came around in the 90ies with many applications maturing only now, electron vortices are new and the field is still rapidly expanding. As with all technique developments in TEM ,the field will only flourish when it provides unique and valuable information in a user friendly way.
Perhaps more important than electron vortices is the paradigm shift from making small electron probes towards optimising the wave function in order to obtain maximum selectivity on a given property of the sample. In other words: what wave function will lead to the best answer to a specific materials science question? In this respect we profit from the increased flexibility that e.g. an aberration corrector offers, but compared to the programmable phase plates that are common in light optics, there is still ample room for improving this flexibility in the future. 





\bibliographystyle{unsrtnat}
\bibliography{chapt_verbeeck}

\begin{thebibliography}{71}
\providecommand{\natexlab}[1]{#1}
\providecommand{\url}[1]{\texttt{#1}}
\expandafter\ifx\csname urlstyle\endcsname\relax
  \providecommand{\doi}[1]{doi: #1}\else
  \providecommand{\doi}{doi: \begingroup \urlstyle{rm}\Url}\fi

\bibitem[Bliokh et~al.(2007)Bliokh, Bliokh, Savel'ev, and Nori]{Bliokh2007}
Konstantin~Yu. Bliokh, Yury~P. Bliokh, Sergey Savel'ev, and Franco Nori.
\newblock Semiclassical dynamics of electron wave packet states with phase
  vortices.
\newblock \emph{Phys. Rev. Lett.}, 99:\penalty0 190404, Nov 2007.
\newblock \doi{10.1103/PhysRevLett.99.190404}.
\newblock URL \url{http://link.aps.org/doi/10.1103/PhysRevLett.99.190404}.

\bibitem[Uchida and Tonomura(2010)]{Uchida2010}
Masaya Uchida and Akira Tonomura.
\newblock {Generation of electron beams carrying orbital angular momentum}.
\newblock \emph{Nature}, 464\penalty0 (7289):\penalty0 737--739, April 2010.
\newblock ISSN 0028-0836.
\newblock \doi{10.1038/nature08904}.
\newblock URL
  \url{http://www.nature.com/nature/journal/v464/n7289/abs/nature08904.html}.

\bibitem[Verbeeck et~al.(2010)Verbeeck, Tian, and
  Schattschneider]{Verbeeck2010}
J.~Verbeeck, H.~Tian, and P.~Schattschneider.
\newblock {Production and application of electron vortex beams.}
\newblock \emph{Nature}, 467\penalty0 (7313):\penalty0 301--4, September 2010.
\newblock ISSN 1476-4687.
\newblock \doi{10.1038/nature09366}.
\newblock URL \url{http://www.ncbi.nlm.nih.gov/pubmed/20844532}.

\bibitem[Schattschneider and Verbeeck(2011)]{SchattschneiderTheoryOf}
P.~Schattschneider and J.~Verbeeck.
\newblock Theory of free electron vortices.
\newblock \emph{Ultramicroscopy}, 111\penalty0 (9–10):\penalty0 1461 -- 1468,
  2011.
\newblock ISSN 0304-3991.
\newblock \doi{10.1016/j.ultramic.2011.07.004}.
\newblock URL
  \url{http://www.sciencedirect.com/science/article/pii/S0304399111001811}.

\bibitem[Saffman(1992)]{saffmanvortex}
P.~G. Saffman.
\newblock \emph{Vortex Dynamics}.
\newblock Cambridge Monographs on Mechanics and Applied Mathematics. Cambridge
  University Press, 1992.
\newblock ISBN 9780521420587.
\newblock URL \url{http://books.google.be/books?id=ksktDE0yVzcC}.

\bibitem[Brandt(2002)]{superconductorvortices}
Ernst~Helmut Brandt.
\newblock Vortices in superconductors.
\newblock \emph{Physica C: Superconductivity}, 369\penalty0 (1–4):\penalty0
  10 -- 20, 2002.
\newblock ISSN 0921-4534.
\newblock \doi{10.1016/S0921-4534(01)01215-1}.
\newblock URL
  \url{http://www.sciencedirect.com/science/article/pii/S0921453401012151}.

\bibitem[Whewell(1833)]{Whewell}
William Whewell.
\newblock Essay towards a first approximation to a map of cotidal lines.
\newblock \emph{Philosophical Transactions of the Royal Society of London},
  123:\penalty0 147--236, 1833.
\newblock URL
  \url{http://rstl.royalsocietypublishing.org/content/123/147.full.pdf+html}.

\bibitem[Nye et~al.(1988)Nye, Hajnal, and Hannay]{nyetides}
J.~F. Nye, J.~V. Hajnal, and J.~H. Hannay.
\newblock Phase saddles and dislocations in two-dimensional waves such as the
  tides.
\newblock \emph{Proceedings of the Royal Society of London. A. Mathematical and
  Physical Sciences}, 417\penalty0 (1852):\penalty0 7--20, 1988.
\newblock \doi{10.1098/rspa.1988.0047}.
\newblock URL
  \url{http://rspa.royalsocietypublishing.org/content/417/1852/7.abstract}.

\bibitem[Silverberg et~al.(2013)Silverberg, Bierbaum, Sethna, and
  Cohen]{moshpitvortex}
Jesse~L. Silverberg, Matthew Bierbaum, James~P. Sethna, and Itai Cohen.
\newblock Collective motion of humans in mosh and circle pits at heavy metal
  concerts.
\newblock \emph{Phys. Rev. Lett.}, 110:\penalty0 228701, May 2013.
\newblock \doi{10.1103/PhysRevLett.110.228701}.
\newblock URL \url{http://link.aps.org/doi/10.1103/PhysRevLett.110.228701}.

\bibitem[Nye and Berry(1974)]{Nye1974}
J.~F. Nye and M.~V. Berry.
\newblock Dislocations in wave trains.
\newblock \emph{Proceedings of the Royal Society of London. A. Mathematical and
  Physical Sciences}, 336\penalty0 (1605):\penalty0 165--190, 1974.
\newblock \doi{10.1098/rspa.1974.0012}.
\newblock URL
  \url{http://rspa.royalsocietypublishing.org/content/336/1605/165.abstract}.

\bibitem[Vaughan and Willetts(1983)]{VaughanWilletts}
J.~M. Vaughan and D.~V. Willetts.
\newblock {Temporal and interference fringe analysis of $TEM^*_{01}$ laser
  modes}.
\newblock \emph{J. Opt. Soc. Am.}, 73\penalty0 (8):\penalty0 1018--1021, Aug
  1983.
\newblock \doi{10.1364/JOSA.73.001018}.
\newblock URL
  \url{http://www.opticsinfobase.org/abstract.cfm?URI=josa-73-8-1018}.

\bibitem[Bazhenov et~al.(1992)Bazhenov, Soskin, and Vasnetsov]{bazhenovscrew}
V.~Yu. Bazhenov, M.~S. Soskin, and M.~V. Vasnetsov.
\newblock Screw dislocations in light wavefronts.
\newblock \emph{Journal of Modern Optics}, 39\penalty0 (5):\penalty0 985--990,
  1992.
\newblock \doi{10.1080/09500349214551011}.
\newblock URL
  \url{http://www.tandfonline.com/doi/abs/10.1080/09500349214551011}.

\bibitem[Lubk et~al.(2013)Lubk, Clark, Guzzinati, and Verbeeck]{lubk2013}
Axel Lubk, Laura Clark, Giulio Guzzinati, and Jo~Verbeeck.
\newblock Topological analysis of paraxially scattered electron vortex beams.
\newblock \emph{Phys. Rev. A}, 87:\penalty0 033834, Mar 2013.
\newblock \doi{10.1103/PhysRevA.87.033834}.
\newblock URL \url{http://link.aps.org/doi/10.1103/PhysRevA.87.033834}.

\bibitem[Allen et~al.(1992)Allen, Beijersbergen, Spreeuw, and
  Woerdman]{Allen1992}
L.~Allen, M.~W. Beijersbergen, R.~J.~C. Spreeuw, and J.~P. Woerdman.
\newblock Orbital angular momentum of light and the transformation of
  {L}aguerre-{G}aussian laser modes.
\newblock \emph{Phys. Rev. A}, 45:\penalty0 8185--8189, Jun 1992.
\newblock \doi{10.1103/PhysRevA.45.8185}.
\newblock URL \url{http://link.aps.org/doi/10.1103/PhysRevA.45.8185}.

\bibitem[Allen et~al.(2003)Allen, Barnett, and Padgett]{OptAngMomBook}
L.~Allen, S.~M. Barnett, and M.~J. Padgett, editors.
\newblock \emph{Optical angular momentum}.
\newblock Institute of Physics Publishing, 2003.
\newblock ISBN 9780750309011.
\newblock URL \url{http://books.google.be/books?id=Vf32PZXJ2gMC}.

\bibitem[Andrews and Babiker(2012)]{BabikerNewBook}
D.L. Andrews and M.~Babiker, editors.
\newblock \emph{The Angular Momentum of Light}.
\newblock Cambridge University Press, 2012.
\newblock ISBN 9781107006348.
\newblock URL \url{http://books.google.be/books?id=li2bysEqHf0C}.

\bibitem[Agyei(2013)]{fusilliphoto}
Nana~B. Agyei.
\newblock Flickr -- {N}anagyei, 2013.
\newblock URL \url{http://www.flickr.com/photos/nanagyei/6655488913/}.
\newblock [Online; accessed 28-June-2013].

\bibitem[Berry and Dennis(2001)]{berry2001knotted}
M.~V. Berry and M.~R. Dennis.
\newblock Knotted and linked phase singularities in monochromatic waves.
\newblock \emph{Proceedings of the Royal Society of London. Series A:
  Mathematical, Physical and Engineering Sciences}, 457\penalty0
  (2013):\penalty0 2251--2263, 2001.
\newblock \doi{10.1098/rspa.2001.0826}.
\newblock URL
  \url{http://rspa.royalsocietypublishing.org/content/457/2013/2251.abstract}.

\bibitem[Dennis(2001)]{dennisthesis}
M.~R. Dennis.
\newblock \emph{Topological singularities in wave fields}.
\newblock PhD thesis, University of Bristol, 2001.
\newblock URL
  \url{http://www.bris.ac.uk/physics/media/theory-theses/dennis-mr-thesis.pdf}.

\bibitem[Andrews(2011)]{andrewsbook}
D.L. Andrews, editor.
\newblock \emph{Structured Light and Its Applications: An Introduction to
  Phase-Structured Beams and Nanoscale Optical Forces}.
\newblock Elsevier Science, 2011.
\newblock ISBN 9780080559667.
\newblock URL \url{http://books.google.be/books?id=BNdCkCXOXX4C}.

\bibitem[Verbeeck et~al.(2013)Verbeeck, Tian, and {Van Tendeloo}]{Verbeeck2013}
Jo~Verbeeck, He~Tian, and Gustaaf {Van Tendeloo}.
\newblock {How to manipulate nanoparticles with an electron beam?}
\newblock \emph{Advanced materials}, 25\penalty0 (8):\penalty0 1114--7,
  February 2013.
\newblock ISSN 1521-4095.
\newblock \doi{10.1002/adma.201204206}.
\newblock URL \url{http://www.ncbi.nlm.nih.gov/pubmed/23184603}.

\bibitem[Peele et~al.(2002)Peele, McMahon, Paterson, Tran, Mancuso, Nugent,
  Hayes, Harvey, Lai, and McNulty]{xrayvortex}
Andrew~G. Peele, Philip~J. McMahon, David Paterson, Chanh~Q. Tran, Adrian~P.
  Mancuso, Keith~A. Nugent, Jason~P. Hayes, Erol Harvey, Barry Lai, and Ian
  McNulty.
\newblock Observation of an x-ray vortex.
\newblock \emph{Opt. Lett.}, 27\penalty0 (20):\penalty0 1752--1754, Oct 2002.
\newblock \doi{10.1364/OL.27.001752}.
\newblock URL \url{http://www.opticsinfobase.org/ol/abstract.cfm?id=70214}.

\bibitem[Thid\'e et~al.(2007)Thid\'e, Then, Sj\"oholm, Palmer, Bergman,
  Carozzi, Istomin, Ibragimov, and Khamitova]{radiovortex}
B.~Thid\'e, H.~Then, J.~Sj\"oholm, K.~Palmer, J.~Bergman, T.~D. Carozzi, Ya.~N.
  Istomin, N.~H. Ibragimov, and R.~Khamitova.
\newblock Utilization of photon orbital angular momentum in the low-frequency
  radio domain.
\newblock \emph{Phys. Rev. Lett.}, 99:\penalty0 087701, Aug 2007.
\newblock \doi{10.1103/PhysRevLett.99.087701}.
\newblock URL \url{http://prl.aps.org/abstract/PRL/v99/i8/e087701}.

\bibitem[Miyamoto et~al.(2011)Miyamoto, Miyagi, Yamada, Furuki, Aoki, Okida,
  and Omatsu]{irvortex}
Katsuhiko Miyamoto, Sachio Miyagi, Masaki Yamada, Kenji Furuki, Nobuyuki Aoki,
  Masahito Okida, and Takashige Omatsu.
\newblock Optical vortex pumped mid-infrared optical parametric oscillator.
\newblock \emph{Opt. Express}, 19\penalty0 (13):\penalty0 12220--12226, Jun
  2011.
\newblock \doi{10.1364/OE.19.012220}.
\newblock URL
  \url{http://www.opticsinfobase.org/oe/abstract.cfm?uri=oe-19-13-12220}.

\bibitem[Terhalle et~al.(2011)Terhalle, Langner, P\"{a}iv\"{a}nranta, Guzenko,
  David, and Ekinci]{uvvortex}
Bernd Terhalle, Andreas Langner, Birgit P\"{a}iv\"{a}nranta, Vitaliy~A.
  Guzenko, Christian David, and Yasin Ekinci.
\newblock Generation of extreme ultraviolet vortex beams using computer
  generated holograms.
\newblock \emph{Opt. Lett.}, 36\penalty0 (21):\penalty0 4143--4145, Nov 2011.
\newblock \doi{10.1364/OL.36.004143}.
\newblock URL
  \url{http://www.opticsinfobase.org/ol/abstract.cfm?uri=ol-36-21-4143}.

\bibitem[Skeldon et~al.(2008)Skeldon, Wilson, Edgar, and
  Padgett]{skeldonacoustic}
K.~D. Skeldon, C.~Wilson, M.~Edgar, and M.~J. Padgett.
\newblock An acoustic spanner and its associated rotational doppler shift.
\newblock \emph{New Journal of Physics}, 10\penalty0 (1):\penalty0 013018,
  2008.
\newblock \doi{10.1088/1367-2630/10/1/013018}.
\newblock URL \url{http://stacks.iop.org/1367-2630/10/i=1/a=013018}.

\bibitem[Torabi and Rezaei(2013)]{torabiacoustic}
Reza Torabi and Zahra Rezaei.
\newblock The effect of {D}irac phase on acoustic vortex in media with screw
  dislocation.
\newblock \emph{Physics Letters A}, 377\penalty0 (28–30):\penalty0 1668 --
  1671, 2013.
\newblock ISSN 0375-9601.
\newblock \doi{10.1016/j.physleta.2013.05.014}.
\newblock URL
  \url{http://www.sciencedirect.com/science/article/pii/S0375960113004842}.

\bibitem[Berkhout and Beijersbergen(2008)]{berkhoutastro}
Gregorius C.~G. Berkhout and Marco~W. Beijersbergen.
\newblock Method for probing the orbital angular momentum of optical vortices
  in electromagnetic waves from astronomical objects.
\newblock \emph{Phys. Rev. Lett.}, 101:\penalty0 100801, Sep 2008.
\newblock \doi{10.1103/PhysRevLett.101.100801}.
\newblock URL \url{http://link.aps.org/doi/10.1103/PhysRevLett.101.100801}.

\bibitem[Wang et~al.(2012)Wang, Yang, Fazal, Ahmed, Yan, Huang, Ren, Yue,
  Dolinar, Tur, and Willner]{wangterabit}
Jian Wang, Jeng-Yuan Yang, Irfan~M Fazal, Nisar Ahmed, Yan Yan, Hao Huang,
  Yongxiong Ren, Yang Yue, Samuel Dolinar, Moshe Tur, and Alan~E. Willner.
\newblock Terabit free-space data transmission employing orbital angular
  momentum multiplexing.
\newblock \emph{Nature Photonics}, 2012.
\newblock \doi{10.1038/nphoton.2012.138}.
\newblock URL
  \url{http://www.nature.com/nphoton/journal/v6/n7/full/nphoton.2012.138.html}.

\bibitem[Roychowdhury et~al.(2004)Roychowdhury, Jaiswal, and Singh]{vortexnot}
Sanjoy Roychowdhury, Virendra~K. Jaiswal, and R.~P. Singh.
\newblock Implementing controlled {NOT} gate with optical vortex.
\newblock \emph{Optics Communications}, 236\penalty0 (4–6):\penalty0 419 --
  424, 2004.
\newblock ISSN 0030-4018.
\newblock \doi{10.1016/j.optcom.2004.03.036}.
\newblock URL
  \url{http://www.sciencedirect.com/science/article/pii/S0030401804003013}.

\bibitem[Durnin et~al.(1987)Durnin, Miceli, and Eberly]{Durnin_bessel_beams}
J.~Durnin, J.~J. Miceli, and J.~H. Eberly.
\newblock Diffraction-free beams.
\newblock \emph{Phys. Rev. Lett.}, 58:\penalty0 1499--1501, Apr 1987.
\newblock \doi{10.1103/PhysRevLett.58.1499}.
\newblock URL \url{http://prl.aps.org/abstract/PRL/v58/i15/p1499_1}.

\bibitem[Verbeeck et~al.(2011{\natexlab{a}})Verbeeck, Schattschneider, Lazar,
  Stoger-Pollach, L\"{o}ffler, Steiger-Thirsfeld, and
  Tendeloo]{Verbeeck_nanoresearch}
J.~Verbeeck, P.~Schattschneider, S.~Lazar, M.~Stoger-Pollach, S.~L\"{o}ffler,
  A.~Steiger-Thirsfeld, and G.~Van Tendeloo.
\newblock Atomic scale electron vortices for nanoresearch.
\newblock \emph{Applied Physics Letters}, 99\penalty0 (20):\penalty0 203109,
  2011{\natexlab{a}}.
\newblock \doi{10.1063/1.3662012}.
\newblock URL \url{http://dx.doi.org/10.1063/1.3662012}.

\bibitem[Bliokh et~al.(2012)Bliokh, Schattschneider, Verbeeck, and
  Nori]{Bliokh2012e}
Konstantin~Y. Bliokh, Peter Schattschneider, Jo~Verbeeck, and Franco Nori.
\newblock {Electron Vortex Beams in a Magnetic Field: A New Twist on Landau
  Levels and Aharonov-Bohm States}.
\newblock \emph{Physical Review X}, 2\penalty0 (4):\penalty0 041011, November
  2012.
\newblock ISSN 2160-3308.
\newblock \doi{10.1103/PhysRevX.2.041011}.
\newblock URL \url{http://link.aps.org/doi/10.1103/PhysRevX.2.041011}.

\bibitem[Bliokh et~al.(2011)Bliokh, Dennis, and Nori]{Bliokh_spin_orbit}
Konstantin~Y. Bliokh, Mark~R. Dennis, and Franco Nori.
\newblock Relativistic electron vortex beams: Angular momentum and spin-orbit
  interaction.
\newblock \emph{Phys. Rev. Lett.}, 107:\penalty0 174802, Oct 2011.
\newblock \doi{10.1103/PhysRevLett.107.174802}.
\newblock URL \url{http://link.aps.org/doi/10.1103/PhysRevLett.107.174802}.

\bibitem[Boxem et~al.(2013)Boxem, Verbeeck, and Partoens]{VanBoxem2013}
Ruben~Van Boxem, Jo~Verbeeck, and Bart Partoens.
\newblock Spin effects in electron vortex states.
\newblock \emph{EPL (Europhysics Letters)}, 102\penalty0 (4):\penalty0 40010,
  2013.
\newblock \doi{10.1209/0295-5075/102/40010}.
\newblock URL \url{http://stacks.iop.org/0295-5075/102/i=4/a=40010}.

\bibitem[Xin and Zheng(2012)]{Xin}
Huolin~L. Xin and Haimei Zheng.
\newblock On-column 2p bound state with topological charge $\pm 1$ excited by
  an atomic-size vortex beam in an aberration-corrected scanning transmission
  electron microscope.
\newblock \emph{Microscopy and Microanalysis}, 18:\penalty0 711--719, 7 2012.
\newblock ISSN 1435-8115.
\newblock \doi{10.1017/S1431927612000499}.

\bibitem[McMorran et~al.(2011)McMorran, Agrawal, Anderson, Herzing, Lezec,
  McClelland, and Unguris]{McMorran2011}
Benjamin~J. McMorran, Amit Agrawal, Ian~M. Anderson, Andrew~A. Herzing,
  Henri~J. Lezec, Jabez~J. McClelland, and John Unguris.
\newblock {Electron vortex beams with high quanta of orbital angular momentum.}
\newblock \emph{Science (New York, N.Y.)}, 331\penalty0 (6014):\penalty0
  192--5, January 2011.
\newblock ISSN 1095-9203.
\newblock \doi{10.1126/science.1198804}.
\newblock URL \url{http://www.sciencemag.org/content/331/6014/192.abstract}.

\bibitem[Berry(2009)]{berryOpticalCurrents}
M.~V. Berry.
\newblock Optical currents.
\newblock \emph{Journal of Optics A: Pure and Applied Optics}, 11\penalty0
  (9):\penalty0 094001, 2009.
\newblock \doi{10.1088/1464-4258/11/9/094001}.
\newblock URL \url{http://stacks.iop.org/1464-4258/11/i=9/a=094001}.

\bibitem[Berkhout(2011)]{berkhoutthesis}
G.~C.~G. Berkhout.
\newblock \emph{Fundamental methods to measure the orbital angular momentum of
  light}.
\newblock PhD thesis, Leiden University, 2011.
\newblock URL \url{https://openaccess.leidenuniv.nl/handle/1887/17842}.

\bibitem[Vasnetsov et~al.(2005)Vasnetsov, Pas'ko, and Soskin]{Vasnetsov}
M.~V. Vasnetsov, V.~A. Pas'ko, and M.~S. Soskin.
\newblock Analysis of orbital angular momentum of a misaligned optical beam.
\newblock \emph{New Journal of Physics}, 7\penalty0 (1):\penalty0 46, 2005.
\newblock \doi{10.1088/1367-2630/7/1/046}.
\newblock URL \url{http://iopscience.iop.org/1367-2630/7/1/046}.

\bibitem[L{\"{o}}ffler and Schattschneider(2012)]{Loffler_crystal}
Stefan L{\"{o}}ffler and Peter Schattschneider.
\newblock {Elastic propagation of fast electron vortices through crystals}.
\newblock \emph{Acta Crystallographica Section A}, 68\penalty0 (4):\penalty0
  443--447, Jul 2012.
\newblock \doi{10.1107/S0108767312013189}.
\newblock URL \url{http://scripts.iucr.org/cgi-bin/paper?S0108767312013189}.

\bibitem[Ivanov and Serbo(2011)]{Ivanov_orbital_helicity}
I.~P. Ivanov and V.~G. Serbo.
\newblock Scattering of twisted particles: Extension to wave packets and
  orbital helicity.
\newblock \emph{Phys. Rev. A}, 84:\penalty0 033804, Sep 2011.
\newblock \doi{10.1103/PhysRevA.84.033804}.
\newblock URL \url{http://pra.aps.org/abstract/PRA/v84/i3/e033804}.

\bibitem[Lloyd et~al.(2012)Lloyd, Babiker, Yuan, and
  Kerr-Edwards]{Lloyd_fields}
S.~M. Lloyd, M.~Babiker, J.~Yuan, and C.~Kerr-Edwards.
\newblock Electromagnetic vortex fields, spin, and spin-orbit interactions in
  electron vortices.
\newblock \emph{Phys. Rev. Lett.}, 109:\penalty0 254801, Dec 2012.
\newblock \doi{10.1103/PhysRevLett.109.254801}.
\newblock URL \url{http://prl.aps.org/abstract/PRL/v109/i25/e254801}.

\bibitem[Voloch-Bloch et~al.(2013)Voloch-Bloch, Lereah, Lilach, Gover, and
  Arie]{Voloch-Bloch2013}
Noa Voloch-Bloch, Yossi Lereah, Yigal Lilach, Avraham Gover, and Ady Arie.
\newblock {Generation of electron Airy beams.}
\newblock \emph{Nature}, 494\penalty0 (7437):\penalty0 331--5, February 2013.
\newblock ISSN 1476-4687.
\newblock \doi{10.1038/nature11840}.
\newblock URL \url{http://www.ncbi.nlm.nih.gov/pubmed/23426323}.

\bibitem[Schattschneider et~al.(2012{\natexlab{a}})Schattschneider,
  St\"{o}ger-Pollach, and Verbeeck]{Schattschneider2012a}
Peter Schattschneider, M.~St\"{o}ger-Pollach, and J.~Verbeeck.
\newblock {Novel Vortex Generator and Mode Converter for Electron Beams}.
\newblock \emph{Physical Review Letters}, 109\penalty0 (8):\penalty0 1--5,
  August 2012{\natexlab{a}}.
\newblock ISSN 0031-9007.
\newblock \doi{10.1103/PhysRevLett.109.084801}.
\newblock URL \url{http://link.aps.org/doi/10.1103/PhysRevLett.109.084801}.

\bibitem[Verbeeck et~al.(2011{\natexlab{b}})Verbeeck, Tian, and
  B\'{e}ch\'{e}]{Verbeeck2011a}
Jo~Verbeeck, He~Tian, and Armand B\'{e}ch\'{e}.
\newblock {A new way of producing electron vortex probes for STEM}.
\newblock \emph{Ultramicroscopy}, 113:\penalty0 83--87, October
  2011{\natexlab{b}}.
\newblock ISSN 03043991.
\newblock \doi{10.1016/j.ultramic.2011.10.008}.
\newblock URL
  \url{http://linkinghub.elsevier.com/retrieve/pii/S0304399111002531}.

\bibitem[Greenshields et~al.(2012)Greenshields, Stamps, and
  Franke-Arnold]{Greenshields2012}
Colin Greenshields, Robert~L. Stamps, and Sonja Franke-Arnold.
\newblock {Vacuum Faraday effect for electrons}.
\newblock \emph{New Journal of Physics}, 14\penalty0 (10):\penalty0 103040,
  October 2012.
\newblock ISSN 1367-2630.
\newblock \doi{10.1088/1367-2630/14/10/103040}.
\newblock URL \url{http://stacks.iop.org/1367-2630/14/i=10/a=103040}.

\bibitem[Guzzinati et~al.(2013)Guzzinati, Schattschneider, Bliokh, Nori, and
  Verbeeck]{Guzzinati2013}
Giulio Guzzinati, Peter Schattschneider, Konstantin~Y. Bliokh, Franco Nori, and
  Jo~Verbeeck.
\newblock {Observation of the Larmor and Gouy Rotations with Electron Vortex
  Beams}.
\newblock \emph{Physical Review Letters}, 110\penalty0 (9):\penalty0 093601,
  February 2013.
\newblock ISSN 0031-9007.
\newblock \doi{10.1103/PhysRevLett.110.093601}.
\newblock URL \url{http://link.aps.org/doi/10.1103/PhysRevLett.110.093601}.

\bibitem[Saitoh et~al.(2012)Saitoh, Hasegawa, Tanaka, and Uchida]{Saitoh2012a}
Koh Saitoh, Yuya Hasegawa, Nobuo Tanaka, and Masaya Uchida.
\newblock {Production of electron vortex beams carrying large orbital angular
  momentum using spiral zone plates.}
\newblock \emph{Journal of electron microscopy}, 61\penalty0 (3):\penalty0
  171--7, June 2012.
\newblock ISSN 1477-9986.
\newblock \doi{10.1093/jmicro/dfs036}.
\newblock URL \url{http://www.ncbi.nlm.nih.gov/pubmed/22394576}.

\bibitem[Danev and Nagayama(2001)]{Danev2001}
R.~Danev and K.~Nagayama.
\newblock {Transmission electron microscopy with Zernike phase plate.}
\newblock \emph{Ultramicroscopy}, 88\penalty0 (4):\penalty0 243--52, September
  2001.
\newblock ISSN 0304-3991.
\newblock URL
  \url{http://linkinghub.elsevier.com/retrieve/pii/S0304399101000882}.

\bibitem[Danev et~al.(2002)Danev, Okawara, Usuda, Kametani, and
  Nagayama]{Danev2002}
R.~Danev, H.~Okawara, N.~Usuda, K.~Kametani, and K.~Nagayama.
\newblock {A novel phase-contrast transmission electron microscopy producing
  high-contrast topographic images of weak objects}.
\newblock \emph{Journal of Biological Physics}, 28\penalty0 (4):\penalty0
  627--635, 2002.
\newblock URL \url{http://www.springerlink.com/index/N5118841393L26XP.pdf}.

\bibitem[Zernike(1942{\natexlab{a}})]{Zernike1942}
F.~Zernike.
\newblock {Phase contrast, a new method for the microscopic observation of
  transparent objects}.
\newblock \emph{Physica}, 9\penalty0 (7):\penalty0 686--698, July
  1942{\natexlab{a}}.
\newblock ISSN 00318914.
\newblock \doi{10.1016/S0031-8914(42)80035-X}.
\newblock URL
  \url{http://linkinghub.elsevier.com/retrieve/pii/S003189144280035X}.

\bibitem[Zernike(1942{\natexlab{b}})]{Zernike1942a}
F.~Zernike.
\newblock {Phase contrast, a new method for the microscopic observation of
  transparent objects part II}.
\newblock \emph{Physica}, 9\penalty0 (10):\penalty0 974--986, December
  1942{\natexlab{b}}.
\newblock ISSN 00318914.
\newblock \doi{10.1016/S0031-8914(42)80079-8}.
\newblock URL
  \url{http://linkinghub.elsevier.com/retrieve/pii/S0031891442800798}.

\bibitem[Schulthei\ss et~al.(2006)Schulthei\ss, Pérez-Willard, Barton,
  Gerthsen, and Schröder]{Schultheiss2006}
K.~Schulthei\ss, F.~Pérez-Willard, B.~Barton, D.~Gerthsen, and R.~R.
  Schröder.
\newblock {Fabrication of a Boersch phase plate for phase contrast imaging in a
  transmission electron microscope}.
\newblock \emph{Review of scientific}, 77\penalty0 (3):\penalty0 90, 2006.
\newblock ISSN 00346748.
\newblock \doi{10.1063/1.2179411}.
\newblock URL \url{http://link.aip.org/link/?RSINAK/77/033701/1
  http://link.aip.org/link/RSINAK/v77/i3/p033701/s1\&Agg=doi}.

\bibitem[Beijersbergen et~al.(1993)Beijersbergen, Allen, {Van der Veen}, and
  Woerdman]{Beijersbergen1993}
M.~W. Beijersbergen, L.~Allen, H.~{Van der Veen}, and J.~P. Woerdman.
\newblock {Astigmatic laser mode converters and transfer of orbital angular
  momentum}.
\newblock \emph{Optics Communications}, 96\penalty0 (1-3):\penalty0 123--132,
  1993.
\newblock URL
  \url{http://www.sciencedirect.com/science/article/pii/003040189390535D}.

\bibitem[Petersen et~al.(2013)Petersen, Weyland, Paganin, Simula, Eastwood, and
  Morgan]{Petersen2013}
T.~C. Petersen, M.~Weyland, D.~M. Paganin, T.~P. Simula, S.~A. Eastwood, and
  M.~J. Morgan.
\newblock {Electron Vortex Production and Control Using Aberration Induced
  Diffraction Catastrophes}.
\newblock \emph{Physical Review Letters}, 110\penalty0 (3):\penalty0 033901,
  January 2013.
\newblock ISSN 0031-9007.
\newblock \doi{10.1103/PhysRevLett.110.033901}.
\newblock URL \url{http://link.aps.org/doi/10.1103/PhysRevLett.110.033901}.

\bibitem[Clark et~al.(2013)Clark, B\'ech\'e, Guzzinati, Lubk, Mazilu,
  Van~Boxem, and Verbeeck]{Clark2013}
L.~Clark, A.~B\'ech\'e, G.~Guzzinati, A.~Lubk, M.~Mazilu, R.~Van~Boxem, and
  J.~Verbeeck.
\newblock Exploiting lens aberrations to create electron-vortex beams.
\newblock \emph{Phys. Rev. Lett.}, 111:\penalty0 064801, Aug 2013.
\newblock \doi{10.1103/PhysRevLett.111.064801}.
\newblock URL \url{http://link.aps.org/doi/10.1103/PhysRevLett.111.064801}.

\bibitem[Tonomura et~al.(1982)Tonomura, Matsuda, Suzuki, Fukuhara, Osakabe,
  Umezaki, Endo, Shinagawa, Sugita, and Fujiwara]{Tonomura1982}
Akira Tonomura, Tsuyoshi Matsuda, Ryo Suzuki, Akira Fukuhara, Nobuyuki Osakabe,
  Hiroshi Umezaki, Junji Endo, Kohsei Shinagawa, Yutaka Sugita, and Hideo
  Fujiwara.
\newblock {Observation of Aharonov-Bohm Effect by Electron Holography}.
\newblock \emph{Physical Review Letters}, 48\penalty0 (21):\penalty0
  1443--1446, May 1982.
\newblock ISSN 0031-9007.
\newblock \doi{10.1103/PhysRevLett.48.1443}.
\newblock URL \url{http://link.aps.org/doi/10.1103/PhysRevLett.48.1443}.

\bibitem[Edgcombe and Loudon(2012)]{Edgcombe2012}
C.~J. Edgcombe and J.~C. Loudon.
\newblock {Use of Aharonov-Bohm effect and chirality control in magnetic phase
  plates for transmission microscopy}.
\newblock \emph{Journal of Physics: Conference Series}, 371:\penalty0 012006,
  July 2012.
\newblock ISSN 1742-6596.
\newblock \doi{10.1088/1742-6596/371/1/012006}.
\newblock URL
  \url{http://stacks.iop.org/1742-6596/371/i=1/a=012006?key=crossref.7ac987ff9728f0920601ae16f557a3ae}.

\bibitem[Tonomura(1987)]{Tonomura1987}
Akira Tonomura.
\newblock {Applications of electron holography}.
\newblock \emph{Reviews of Modern Physics}, 59\penalty0 (3):\penalty0 639--669,
  July 1987.
\newblock ISSN 0034-6861.
\newblock \doi{10.1103/RevModPhys.59.639}.
\newblock URL \url{http://link.aps.org/doi/10.1103/RevModPhys.59.639}.

\bibitem[Schattschneider et~al.(2012{\natexlab{b}})Schattschneider, Schaffer,
  Ennen, and Verbeeck]{Schattschneider2012}
Peter Schattschneider, Bernhard Schaffer, I.~Ennen, and J.~Verbeeck.
\newblock {Mapping spin-polarized transitions with atomic resolution}.
\newblock \emph{Physical Review B}, 85\penalty0 (13):\penalty0 134422, April
  2012{\natexlab{b}}.
\newblock ISSN 1098-0121.
\newblock \doi{10.1103/PhysRevB.85.134422}.
\newblock URL \url{http://link.aps.org/doi/10.1103/PhysRevB.85.134422}.

\bibitem[Sch\"utz et~al.(1987)Sch\"utz, Wagner, Wilhelm, Kienle, Zeller, Frahm,
  and Materlik]{Schutz1987}
G.~Sch\"utz, W.~Wagner, W.~Wilhelm, P.~Kienle, R.~Zeller, R.~Frahm, and
  G.~Materlik.
\newblock Absorption of circularly polarized {X}-rays in iron.
\newblock \emph{Phys. Rev. Lett.}, 58:\penalty0 737--740, Feb 1987.
\newblock \doi{10.1103/PhysRevLett.58.737}.
\newblock URL \url{http://link.aps.org/doi/10.1103/PhysRevLett.58.737}.

\bibitem[Schattschneider et~al.(2006)Schattschneider, Rubino, H\'{e}bert, Rusz,
  Kunes, Nov\'{a}k, Carlino, Fabrizioli, Panaccione, and
  Rossi]{Schattschneider2006}
P.~Schattschneider, S.~Rubino, C.~H\'{e}bert, J.~Rusz, J.~Kunes, P.~Nov\'{a}k,
  E.~Carlino, M.~Fabrizioli, G.~Panaccione, and G.~Rossi.
\newblock {Detection of magnetic circular dichroism using a transmission
  electron microscope.}
\newblock \emph{Nature}, 441\penalty0 (7092):\penalty0 486--8, May 2006.
\newblock ISSN 1476-4687.
\newblock \doi{10.1038/nature04778}.
\newblock URL \url{http://www.ncbi.nlm.nih.gov/pubmed/16724061}.

\bibitem[Schattschneider et~al.(2010)Schattschneider, Ennen, Löffler,
  Stöger-Pollach, and Verbeeck]{Schattschneider2010_with_Ennen}
P.~Schattschneider, I.~Ennen, S.~Löffler, M.~Stöger-Pollach, and
  J.~Verbeeck.
\newblock {Circular dichroism in the electron microscope: Progress and
  applications (invited)}.
\newblock \emph{Journal of Applied Physics}, 107\penalty0 (9):\penalty0 09D311,
  2010.
\newblock ISSN 00218979.
\newblock \doi{10.1063/1.3365517}.
\newblock URL
  \url{http://link.aip.org/link/JAPIAU/v107/i9/p09D311/s1\&Agg=doi}.

\bibitem[Schattschneider et~al.(2012{\natexlab{c}})Schattschneider, Schaffer,
  Ennen, and Verbeeck]{Schattschneider_spin_polarized}
P.~Schattschneider, B.~Schaffer, I.~Ennen, and J.~Verbeeck.
\newblock Mapping spin-polarized transitions with atomic resolution.
\newblock \emph{Phys. Rev. B}, 85:\penalty0 134422, Apr 2012{\natexlab{c}}.
\newblock \doi{10.1103/PhysRevB.85.134422}.

\bibitem[Pettit et~al.(1975)Pettit, Silcox, and Vincent]{Pettit1975}
R.~B. Pettit, J.~Silcox, and R.~Vincent.
\newblock Measurement of surface-plasmon dispersion in oxidized aluminum films.
\newblock \emph{Phys. Rev. B}, 11:\penalty0 3116--3123, Apr 1975.
\newblock \doi{10.1103/PhysRevB.11.3116}.
\newblock URL \url{http://link.aps.org/doi/10.1103/PhysRevB.11.3116}.

\bibitem[Nelayah et~al.(2007)Nelayah, Kociak, St{\'e}phan, de~Abajo, Tenc{\'e},
  Henrard, Taverna, Pastoriza-Santos, Liz-Marz{\'a}n, and Colliex]{Nelayah2007}
Jaysen Nelayah, Mathieu Kociak, Odile St{\'e}phan, F.~Javier~Garcia de~Abajo,
  Marcel Tenc{\'e}, Luc Henrard, Dario Taverna, Isabel Pastoriza-Santos,
  Luis~M. Liz-Marz{\'a}n, and Christian Colliex.
\newblock {Mapping surface plasmons on a single metallic nanoparticle}.
\newblock \emph{Nature Physics}, 3\penalty0 (5):\penalty0 348--353, April 2007.
\newblock ISSN 1745-2473.
\newblock \doi{10.1038/nphys575}.
\newblock URL \url{http://www.nature.com/nphys/journal/v3/n5/pdf/nphys575.pdf}.

\bibitem[Mohammadi et~al.(2012)Mohammadi, {Van Vlack}, Hughes, Bornemann, and
  Gordon]{Mohammadi2012}
Zeinab Mohammadi, Cole~P {Van Vlack}, Stephen Hughes, Jens Bornemann, and
  Reuven Gordon.
\newblock {Vortex electron energy loss spectroscopy for near-field mapping of
  magnetic plasmons.}
\newblock \emph{Optics express}, 20\penalty0 (14):\penalty0 15024--34, July
  2012.
\newblock ISSN 1094-4087.
\newblock URL \url{http://www.ncbi.nlm.nih.gov/pubmed/22772198}.

\bibitem[Smith et~al.(2004)Smith, Pendry, and Wiltshire]{Smith2004}
D.~R. Smith, J.~B. Pendry, and M.~C.~K. Wiltshire.
\newblock Metamaterials and negative refractive index.
\newblock \emph{Science}, 305:\penalty0 788–792, 2004.
\newblock URL \url{http://www.sciencemag.org/content/305/5685/788.short}.

\bibitem[Al\`{u} and Engheta(2007)]{Alu2007}
A.~Al\`{u} and N.~Engheta.
\newblock Cloaking and transparency for collections of particles with
  metamaterial and plasmonic covers.
\newblock \emph{Opt. Express}, 15:\penalty0 7578--7590, 2007.
\newblock \doi{10.1088/1464-4258/10/9/093002}.
\newblock URL \url{http://stacks.iop.org/1464-4258/10/i=9/a=093002}.

\bibitem[Karimi et~al.(2012)Karimi, Marrucci, Grillo, and
  Santamato]{Karimi2012}
Ebrahim Karimi, Lorenzo Marrucci, Vincenzo Grillo, and Enrico Santamato.
\newblock {Spin-to-Orbital Angular Momentum Conversion and Spin-Polarization
  Filtering in Electron Beams}.
\newblock \emph{Physical Review Letters}, 108\penalty0 (4):\penalty0 044801,
  January 2012.
\newblock ISSN 0031-9007.
\newblock \doi{10.1103/PhysRevLett.108.044801}.
\newblock URL \url{http://link.aps.org/doi/10.1103/PhysRevLett.108.044801}.

\end{thebibliography}








\end{document}